\documentclass[twocolumn,prb,showpacs,amsmath,amssymb]{revtex4}
\usepackage{graphicx}
\usepackage{epsfig}
\usepackage{dcolumn}
\usepackage{bm}

\begin{document}
\title{ Possibility of superconductivity in graphite intercalated with
alkaline earths investigated with density functional theory. }
\author{Matteo Calandra}
\address{Institut de Min\'eralogie et de Physique des Milieux condens\'es, 
case 115, 4 place Jussieu, 75252, Paris cedex 05, France}
\author{ Francesco Mauri}
\address{Institut de Min\'eralogie et de Physique des Milieux condens\'es, 
case 115, 4 place Jussieu, 75252, Paris cedex 05, France}
\date{\today}

\begin{abstract}
Using density functional theory we investigate the occurrence of 
superconductivity in AC$_6$ with A=Mg,Ca,Sr,Ba. We predict that at zero pressure,
Ba and Sr should be superconducting with critical temperatures (T$_c$) 
0.2 K and 3.0 K, respectively. We study the pressure dependence 
of T$_c$ assuming the same symmetry for the crystal structures at zero and finite
pressures. 
We find that  the SrC$_6$  and BaC$_6$
critical temperatures should be substantially enhanced by pressure. 
On the contrary, for CaC$_6$ we find that
in the 0 to 5 GPa region, T$_c$ weakly increases with pressure.
The increase is much smaller than what shown in several recent experiments.
Thus we suggest that in CaC$_6$ a continous phase transformation, such as an increase
in staging, occurs at finite pressure.
Finally we argue that, although MgC$_6$ is unstable, the synthesis
of intercalated systems of the kind Mg$_x$Ca$_{1-x}$C$_y$ could lead
to higher critical temperatures.
\end{abstract}
\pacs{63.20.Kr, 63.20.Dj , 78.30.Er, 74.70.Ad}

\maketitle
\section{Introduction}
It has long been known that Graphite intercalated compounds (GICs)
can display a superconducting behavior at low temperature
\cite{DresselhausRev}. However, until the discovery of 
ytterbium and calcium intercalated graphite \cite{Weller,Genevieve}
(T$_c$(YbC$_6$)=6.5 K and T$_c$(CaC$_6$)=11.5 K), 
the critical temperatures achieved 
were typically inferior to 5 Kelvin. Very recently, it has been
shown that even higher critical temperatures (up to 15.1 K)
can be achieved in CaC$_6$ applying hydrostatic
pressure (up to 8 GPa)\cite{Gauzzi}. This is presently the
highest T$_c$ reported in a GIC.

From a practical point of view GICs are appealing since carbon is
a very versatile element. 
Beside the fact that the three most known
carbon crystalline structures (Fullerene, Graphite and Diamond)
display superconducting behavior upon intercalation
\cite{DresselhausRev,OlleRev} or doping\cite{Ekimov},
carbon is also interesting for the possibility of making nanotubes
which can display a superconducting behavior \cite{Tang}.
Moreover the number of reagents  
which can be intercalated in graphite, 
by using chemical methods or high pressure synthesis, is very
large. Consequently finding GICs with even higher critical
temperatures is not a remote possibility.
Therefore theoretical calculations can be a precious tool
for the investigation of superconductivity in GICs.

The nature of superconductivity in intercalated GICs is still controversial.
Due to their layered nature and to the presence of an intercalant band
at the Fermi level (E$_f$), Cs\'anyi {\it et al.} \cite{Csanyi} suggested a
non-conventional exciton pairing mechanism \cite{Allender}. 
On the contrary Mazin \cite{Mazin} proposed an electron-phonon mechanism mainly
sustained by Ca vibration.
In a previous work\cite{Calandra} we suggested that the pairing is 
mediated by the electron-phonon interaction. In particular the carriers
are mostly electrons in the Ca Fermi surfaces coupled with Ca in-plane and
C-out of plane vibrations. The calculation of the isotope effect coefficient
showed that the contribution of Ca in-plane vibrations and C out-of-plane
vibrations to superconductivity is comparable.

Experimental data seem to confirm that the pairing in CaC$_6$ and YbC$_6$
is indeed due to the electron-phonon interaction, but several open
questions remain.
From the exponential behavior of the penetration depth in CaC$_6$
Lamura {\it et al.} \cite{Lamura} deduced an isotropic gap of magnitude 
$\Delta$(0)=1.79$\pm$ 0.08 meV.
A similar isotropic gap ($\Delta$(0)=1.6$\pm$ 0.2 meV) 
has been measured by scanning tunneling 
spectroscopy \cite{Bergeal}. The corresponding values of $2\Delta(0)/T_c$ 
are in agreement with the BCS theory. Similar conclusions have been 
obtained from the specific heat jump at superconducting transition\cite{Kim2}.
Thermal conductivity data in the presence of a magnetic field
 \cite{Sutherland} indicate that in YbC$_6$ the order parameter has s-wave
symmetry too and exclude the occurrence of multiple gaps.
Isotope effect measurements \cite{Hinks} show a huge Ca isotope
coefficient, $\alpha({\rm Ca})=0.5$, in disagreement with theoretical
calculations \cite{Calandra}. This is surprising, since the gap
and specific heat data are correctly described by DFT calculations,
meaning that the calculated electron-phonon coupling is probably correct.
Even more interesting is that 
the total isotope effect would be probably larger than 0.5, although
C-isotope effect measurements are necessary to confirm this and to judge the 
validity of the measurements of $\alpha({\rm Ca})$.
In any case, the large Ca-isotope coefficient, the measured
superconducting gap and the jump of the specific heat at the
transition they all go in favor of a phonon mediated mechanism
with, most likely, a single s-wave gap. 

In this work we want to push a step forward the prediction that can be made
by the electron-phonon theory. This is important since it may allow
to identify GICs with higher critical temperatures and it also represents
a significative benchmark for DFT simulations.
It was noted \cite{Csanyi} that all the 
superconducting GICs possess an intercalant Fermi surfaces at E$_f$. This fact is
relevant for both pairing mechanisms proposed. 
In the case of a conventional electron-phonon mechanism the electrons
in the intercalant Fermi surface are the ones strongly coupled to the phonons
\cite{Calandra}.
Thus we study  the possible occurrence 
of superconductivity in graphite intercalated with alkaline earths
(AC$_6$ with A=Ba,Sr,Mg). All these GICS have an intercalant 
Fermi surface at E$_f$ so they are good candidates for superconductivity.
We predict the critical temperatures for BaC$_6$ and SrC$_6$ in the 
framework of the electron-phonon coupling theory. 

As mentioned before, the critical temperature of CaC$_6$ and YbC$_6$ is
enhanced with pressure.
Resistivity measurements under pressure show that, at $\approx$ 8 GPa,
CaC$_6$ undergoes a structural phase transition to a new superconducting
phase with a lower critical temperature. 
The new structure seems to be stable at least up to 16 GPa. 
In other successive works
 magnetic\cite{Kim1,Smith} and resistive \cite{Smith} measurements 
were performed in a much smaller
range of pressures (0-1.6GPa) and the behavior observed 
was confirmed. Pressure also increases
the critical temperature of YbC$_6$ up to approximatively 7.0 K
at $\approx$2.0 GPa. In this case too a structural transition
is seen towards a new superconducting phase with lower T$_c$.
In both YbC$_6$ and CaC$_6$ the dependence of
T$_c$ is linear with similar $\Delta T_c/\Delta P$ (0.4 for YbC$_6$
and 0.5 for CaC$_6$). 
The fact that T$_c$ can be enhanced with pressure suggests that
this can be a general mechanism for superconducting in GICs.
It is then important to study superconductivity as a function
of pressure in BaC$_6$,SrC$_6$ and finally CaC$_6$.

After illustrating in section \ref{sec:technical} and \ref{sec:structures} 
the technical parameters and the lattice structures used in the
simulations, we study the
electronic structure (sec. \ref{sec:el-structure} ), the phonon dispersions
(sec. \ref{sec:phonon}) and the superconducting properties 
(sec. \ref{sec:SC}) of Alkaline-earth intercalated graphite at ambient pressure
and at finite pressure. Particular emphasis is given to the case of CaC$_6$
under pressure (sec. \ref{sec:CaC6P}).

\section{Technical details}\label{sec:technical}

Density Functional Theory (DFT) calculations are performed using the 
espresso code\cite{PWSCF} and
the generalized gradient approximation
(GGA) \cite{PBE}. We use ultrasoft pseudopotentials\cite{Vanderbilt}
with valence configurations 3s$^2$3p$^6$4s$^2$ for Ca,
4s$^2$4p$^6$4d$^1$5s$^1$5p$^0$ for Sr, 
5s$^2$5p$^6$5d$^0$6s$^2$6p$^0$ for Ba, and 
2s$^2$2p$^2$ for C. For Mg we use Troullier-Martin\cite{Troullier} pseudopotentials
with configuration 3s$^{0.1}$3p$^0$3d$^0$.
The wavefunctions and the charge density are expanded using a 35 Ry
and a 600 Ry cutoff. 
The dynamical matrices and the electron-phonon coupling are calculated using
Density Functional Perturbation Theory in the linear response\cite{PWSCF}.
For the electronic integration in the phonon calculation (structure R\={3}m) 
we use a $N_{k}=6\times6\times6$ and $N_{k}=8\times8\times8$ uniform k-point 
meshes and Hermite-Gaussian smearing ranging from 0.1  to 0.05 Ry.
In order to obtain very accurate phonon-frequencies for the low energy
modes (below 15 meV) it is crucial to use a large cutoff for the 
charge density (600 Ry at least) and a very high convergence
threshold in the phonon calculations.
For the evaluation of the electron-phonon coupling and 
of the electronic density of states
 we use  $N_k=25\times 25\times 25$ and $N_k=20\times 20\times 20$ 
meshes respectively.
For the $\lambda$ average over the phonon momentum {\bf q} 
we use a  $N_q=4\times 4\times 4\,$ ${\bf q}-$points mesh.
The phonon dispersion is obtained by Fourier interpolation of the 
dynamical matrices computed on the $N_q$ mesh. 

\section{Crystal structure}\label{sec:structures}

All the considered compounds are stage 1 \cite{DresselhausRev} at zero pressure.
The atomic structure\cite{Genevieve} of CaC$_{6}$ involves a stacked arrangement 
of graphene sheets (stacking AAA) with Ca atoms occupying interlayer sites above 
the centers of the hexagons (stacking $\alpha\beta\gamma$).  
The crystallographic structure is R\={3}m where 
the Ca atoms occupy the 1a Wyckoff position (0,0,0) and the C atoms 
the 6g positions (x,-x,1/2)
with x$=1/6$. 
SrC$_{6}$ and BaC$_6$ have a slightly different crystal structure
\cite{Guerard} involving a stacked arrangement 
of graphene sheets (stacking AAA) with Sr and Ba atoms occupying 
interlayer sites above the centers of the hexagons with an $\alpha\beta$ stacking. 
The crystallographic structure is P6$_{3}$/mmc where 
the Sr,Ba atoms occupy the 2c Wyckoff position (1/3,2/3,1/4) and (1/3,2/3,3/4)  
and the C atoms 
the 12i positions (1/3,0,0). The experimental in plane lattice parameter a and
the interlayer spacing c for the three
structures are illustrated in table \ref{tab:structures}. 

\begin{table}[hbt]
\begin{ruledtabular}
\begin{tabular}{lccccc} 
Material & Stacking          & a,c           & a,c (LDA)        &     a,c (GGA)        \\
BaC$_6$  & A$\alpha$A$\beta$ & 4.302  5.25   &   4.280  5.00     &  4.350    5.20     \\
SrC$_6$  & A$\alpha$A$\beta$ & 4.315  4.95   &   4.285 4.80     &  4.325   5.00    \\
CaC$_6$  & A$\alpha$A$\beta$A$\gamma$  & 4.333  4.524  &    4.29  4.36    &  4.333    4.51   \\ 
MgC$_6$  &                   &               &  4.317 3.75  & 4.35 3.95 \\ 
\end{tabular}
\end{ruledtabular}
\caption{Experimental (first two columns) structural parameters (Angstrom)
of BaC$_6$, SrC$_6$, CaC$_6$. With $c$ we indicate the interlayer
spacing, namely the distance between two graphene layers, while a
is the in plane lattice parameter.
In the last two  columns we report the parameters of the 
theoretical (LDA and GGA) R\={3}m structure 
(stacking A$\alpha$A$\beta$A$\gamma$)
having the same a and the same interlayer distance of the experimental structure. 
This rhombohedral structure is considered in all the calculations. Since MgC$_6$ has
has never been synthesized we only report its theoretical minimized structure 
with symmetry R\={3}m}
\label{tab:structures}
\end{table}

Even if the structural symmetry of SrC$_6$ and BaC$_6$ are different from the one
of CaC$_6$, in this work we consider the same rhombohedral symmetry group (namely
R\={3}m with stacking A$\alpha$A$\beta$A$\gamma$ ) for all the considered GICs.
We do not expect this assumption to affect qualitatively the 
calculated electronic and phonon properties since the two structures differ only
for large distance neighbors. Indeed the distances of the first and second
nearest neighbors are the same. The differences between the metal lattice sites 
in the A$\alpha$A$\beta$A$\gamma$ and in the A$\alpha$A$\beta$ structures is 
equivalent to those existing in the fcc and in the hcp structures.

In the last three columns of table \ref{tab:structures} we
report the theoretically minimized parameters assuming the R\={3}m structure 
and using the local density approximation (LDA) 
or the Generalized-Gradient approximation (GGA).
The interlayer distance between graphite layers (c) and the in-plane lattice
parameter (a) calculated with GGA are in very good agreement with experiments. 
The equilibrium LDA a and c parameters are slightly compressed respect
to the CGA and experimental values.

Since  MgC$_6$ has never been synthesized, we also assume that it crystallizes in 
the R\={3}m structure. The theoretical equilibrium parameters are reported in table  
\ref{tab:structures}. We have verified that this structure is unstable since its
energy in CGA is lower by 0.016 Ry/(Cell CaC$_6$) then the ones of Magnesium and 
Graphite separated.

In Alkali-earths GICs as the atomic number of the intercalant (Z) is reduced
the c parameter is also reduced. To disentangle the two effects we consider
BaC$_6$, SrC$_6$ and CaC$_6$ under isotropic pressure. 
Since the structure of these systems
at a given pressure is not known, we minimize the R\={3}m structure 
at a given isotropic pressure. 
The lattice parameters obtained are illustrated in
table \ref{tab:structuresP}.
\begin{table}[]
\begin{ruledtabular}
\begin{tabular}{lccc}
System   & Pressure (GPa) & a        &   c      \\ \hline 
 BaC$_6$ &  8    & 4.287   & 4.925 \\ \hline
 SrC$_6$ &  8    & 4.295   & 4.750 \\
         & 16.5  & 4.265   & 4.524 \\ \hline
 CaC$_6$ &  3    & 4.333   & 4.524 \\ 
         &  5    & 4.317   & 4.349 \\
         &  6    & 4.314   & 4.330 \\
         &  7    & 4.310   & 4.301 \\
         &  8    & 4.307   & 4.290 \\
         &  9    & 4.303   & 4.270 \\
         & 10    & 4.300   & 4.250 \\ 
\end{tabular}
\end{ruledtabular}
\caption{Theoretical (GGA) Structure of Alkali-earth GICs under pressure}
\label{tab:structuresP}
\end{table}
It should be noticed that for SrC$_6$ the pressure of 16.5 GPa has been 
chosen since the c parameter has the same value as in CaC$_6$ at zero pressure.

When possible, we always used the experimental a and c parameters in the
electronic structure calculations. When lacking, we used the GGA minimized
parameters.

\section{Electronic Structure}\label{sec:el-structure}
\begin{figure*}[]
{\includegraphics[height=12.0cm]{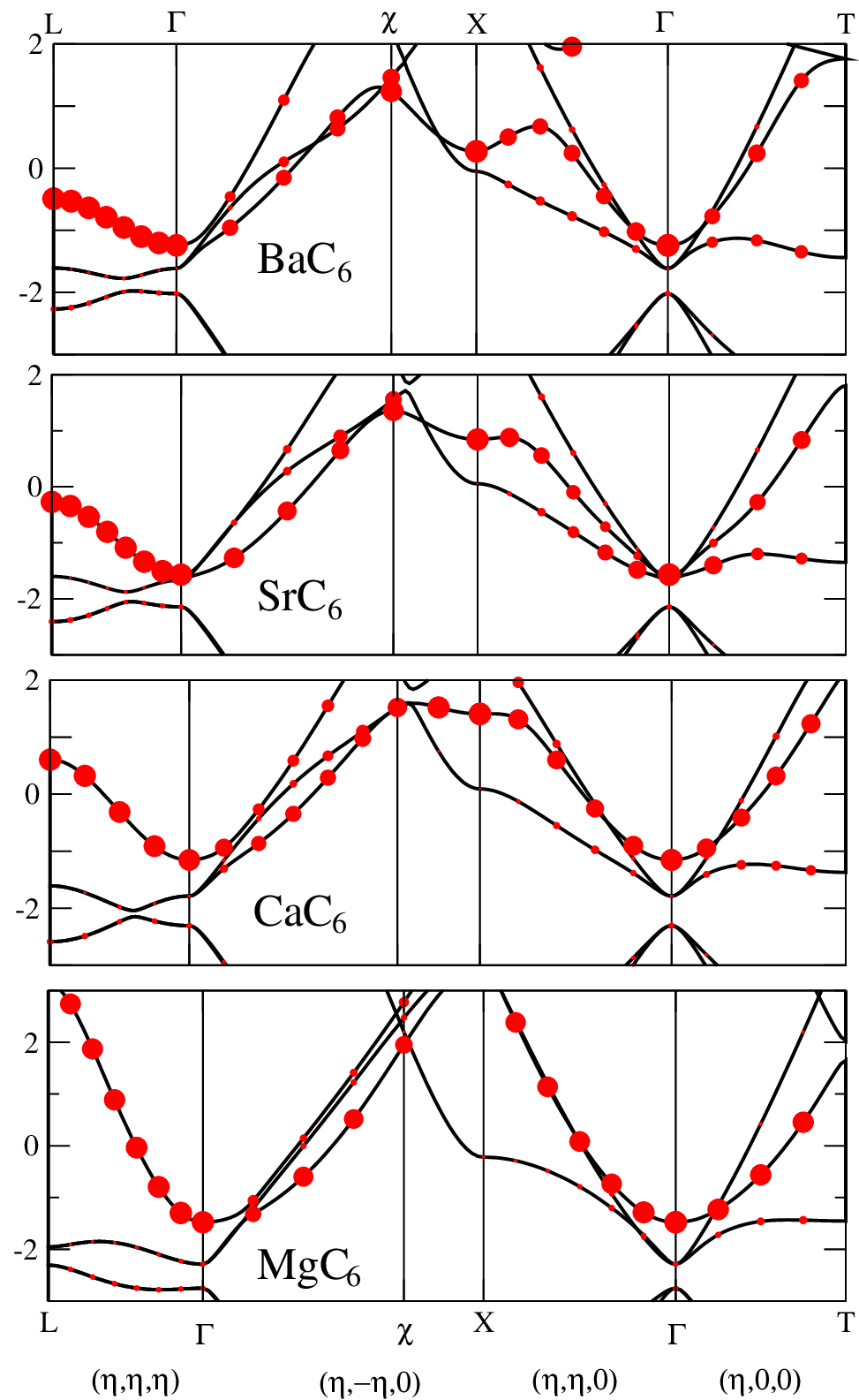}%
\includegraphics[height=12.0cm]{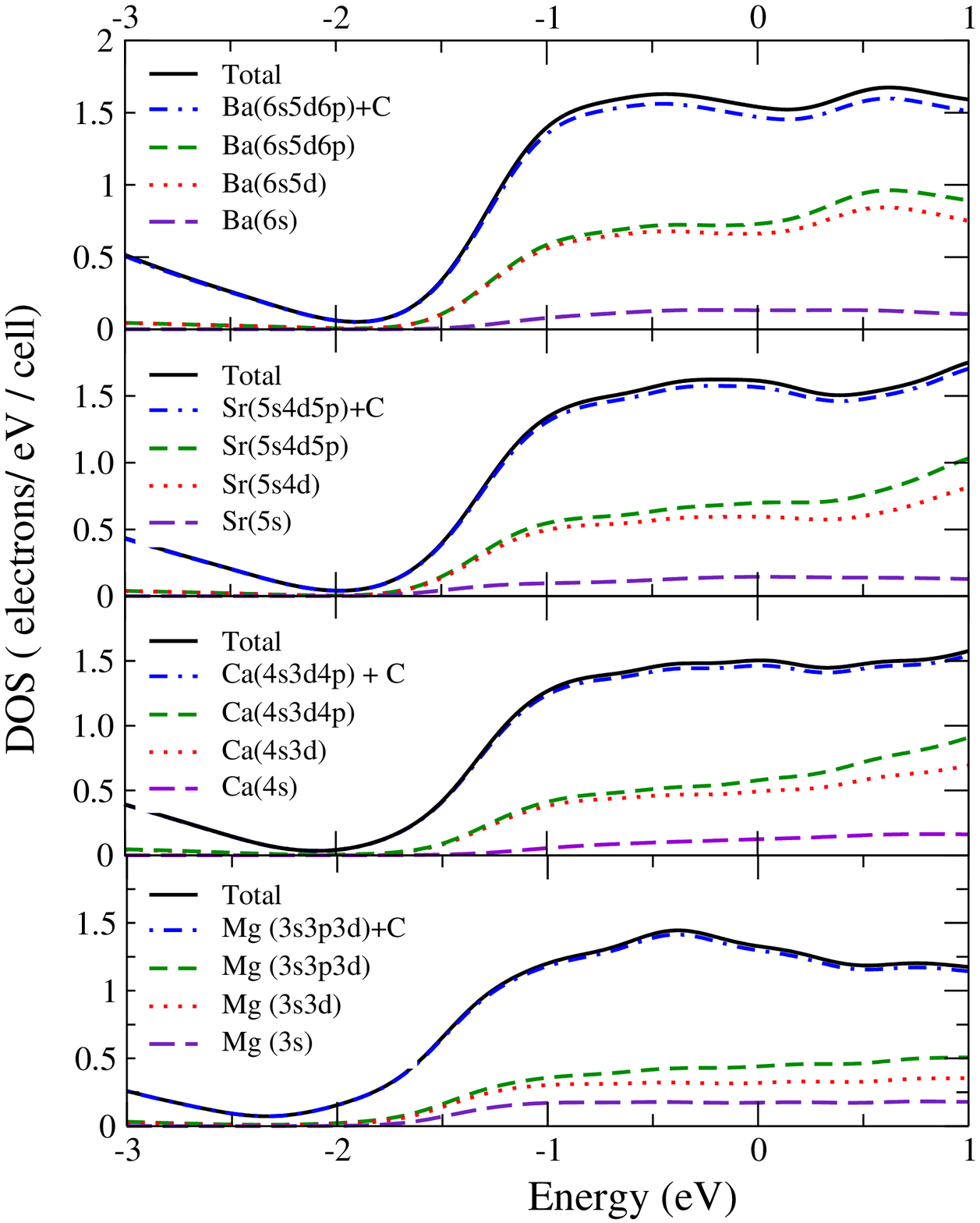}}%
\caption{(Color online) Band structure (left) and density of states projected
over atomic orbitals (right) for several Alkali-earths GICs. The zero energy
corresponds to the Fermi energy. The size of the
dots represents the percentage of intercalant component. As a reference, in CaC$_6$ 
the dot at $\approx 0.6 eV$ at the L point represents 95\%.}
\label{fig:bandsDOSP0}
\end{figure*}

The zero-pressure DFT-band-structures of the Ba,Sr,Ca,Mg intercalated compounds 
are illustrated  in fig. \ref{fig:bandsDOSP0}. All the considered
alkali-earths intercalated GICs have at least one intercalant band at the Fermi level.
This was proposed to be
\cite{Calandra,Csanyi} 
a necessary condition in order to have superconductivity with a reasonable 
critical temperature in GICs.

\begin{figure}[h]
\includegraphics[width=8.0cm]{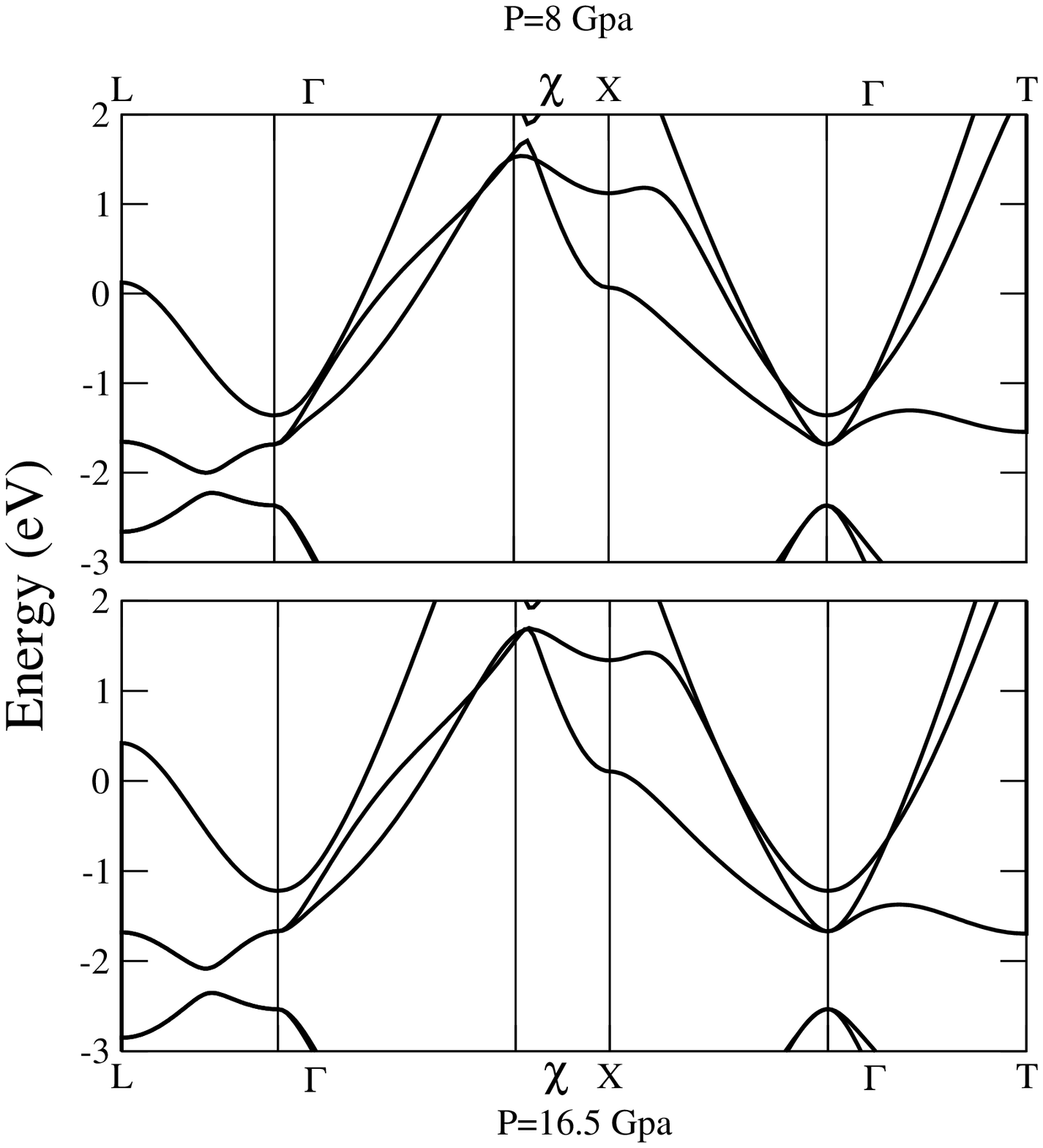}%
\caption{Band structure of SrC$_6$ at different pressures. The zero of the 
energy corresponds to the Fermi energy}
\label{fig:SrC6BandsP}
\end{figure}

The considered GICs are layered structures, but their band structure are 
clearly three-dimensional. This was understood in the case of CaC$_6$\cite{Calandra} 
since the rhombohedral angle (49.55$^o$) is not too far from the 
one corresponding to the fcc structure (60$^o$). The rhombohedral angles for
 BaC$_6$, SrC$_6$ and MgC$_6$ are 43.47$^o$,45.81$^o$, 55.37$^o$, respectively.
Thus the rhombohedral angle is larger as c is reduced.
This is even more evident comparing the behavior of the band-structure as
a function of pressure in fig. \ref{fig:SrC6BandsP} for SrC$_6$. Note
in particular, that at 16.5 GPa (where the c parameter is the same as
that of CaC$_6$ at zero pressure) the band structure is very similar to
the one of CaC$_6$ at zero pressure. 
The larger bandwidths of the intercalant band as c is reduced reflects
the higher three dimensional character of the structure. 

The electronic density of states for the four considered compounds 
are very similar (see fig. \ref{fig:bandsDOSP0}) 
and also their values at the Fermi 
level are very close. 
No qualitative difference in the superconducting behavior of these systems 
can be attributed to the number of carriers at the Fermi energy 
(i.e. to the different value of N(0)). 

The atomic-projected density of states 
is calculated using the L\"owdin populations,
$\rho_{\eta}(\epsilon)=
\frac{1}{N_k}\sum_{{\bf k}n}|\langle \phi^{L}_{\eta}|\psi_{{\bf k}n}\rangle|^2
\delta(\epsilon_{{\bf k}n}-\epsilon)$. 
In this expression
$|\phi^{L}_{\eta}\rangle=
\sum_{\eta\prime}[{\bf S}^{-1/2}]_{\eta,\eta^{\prime}} |\phi^{a}_{\eta^{\prime}
}\rangle$ 
are the orthonormalized L\"owdin orbitals, 
$ |\phi^{a}_{\eta^{\prime}}\rangle$ are the
atomic wavefunctions and
$ S_{\eta,\eta^{\prime}}=\langle \phi^{a}_{\eta} |\phi^{a}_{\eta^{\prime}}\rangle$.
The density of states projected over the intercalant states has very similar value at
E$_f$ for the three case and it is slightly reduced with decreasing Z. No
particular differences respect to the case of CaC$_6$ are observed\cite{Calandra}.
In fig. \ref{fig:bandsDOSP0} (right) the size of the dots represents
the projection over the intercalant atomic states. 

\begin{figure}[h]
{\includegraphics[width=2.75cm]{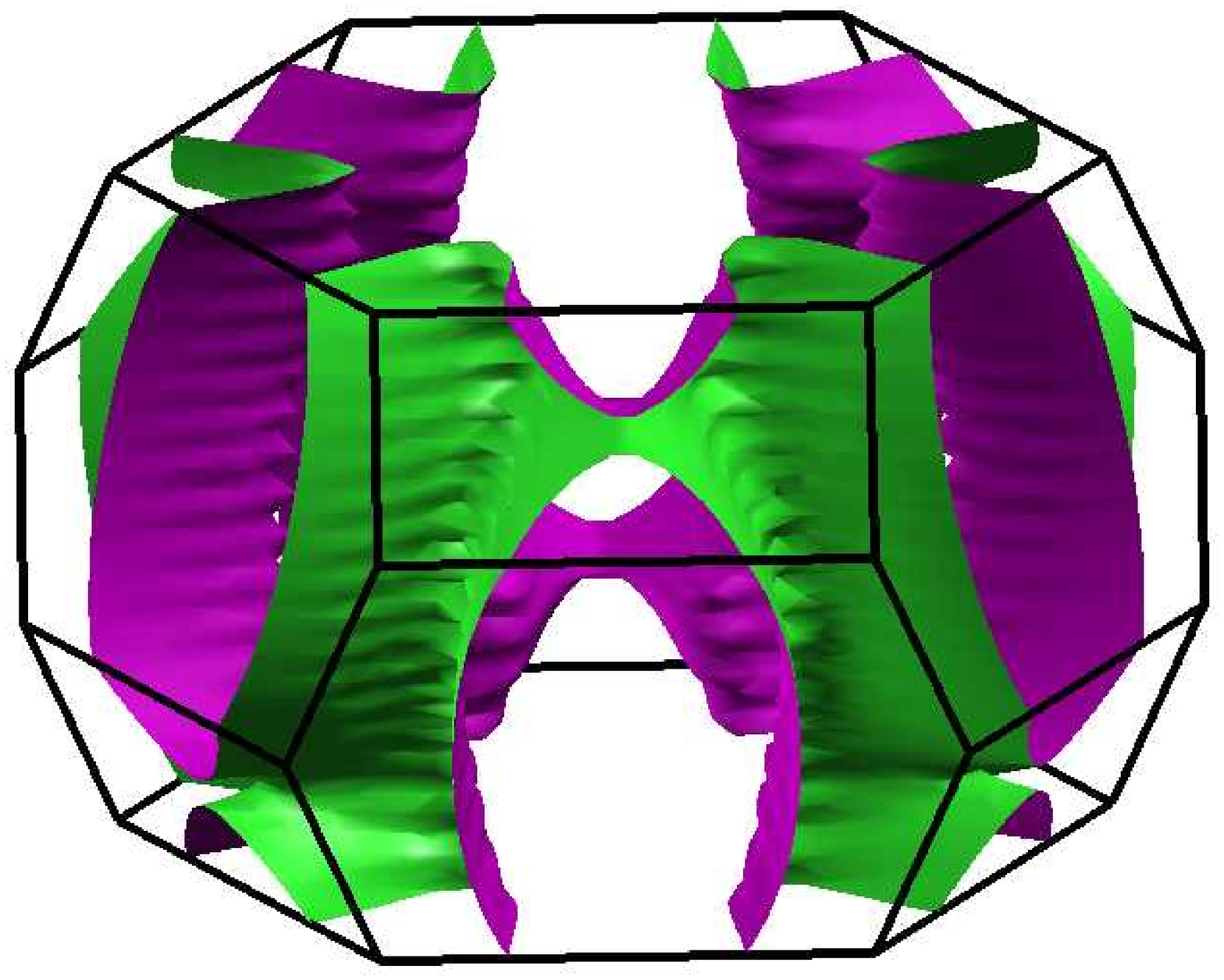}%
\includegraphics[width=2.75cm]{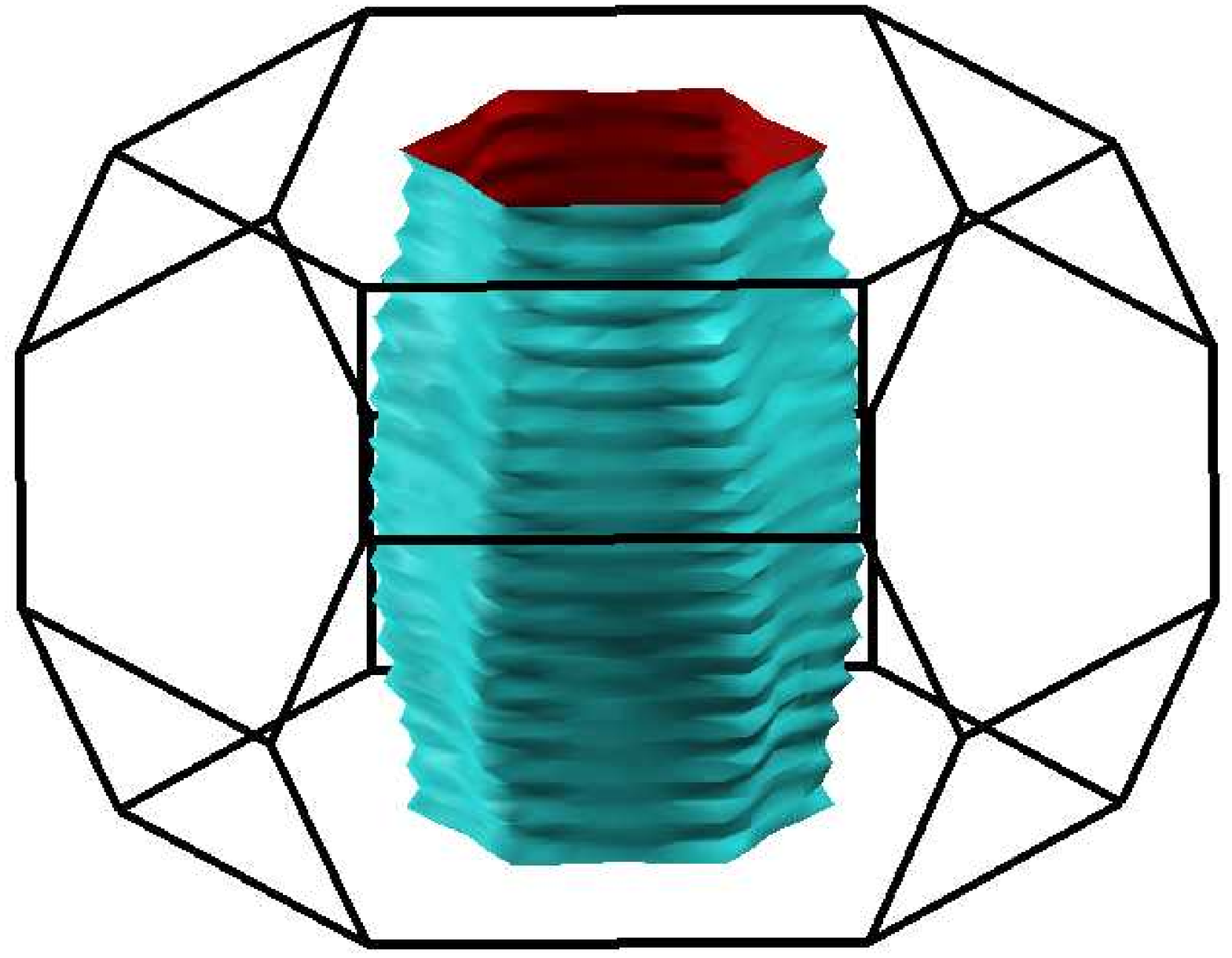}%
\includegraphics[width=2.75cm]{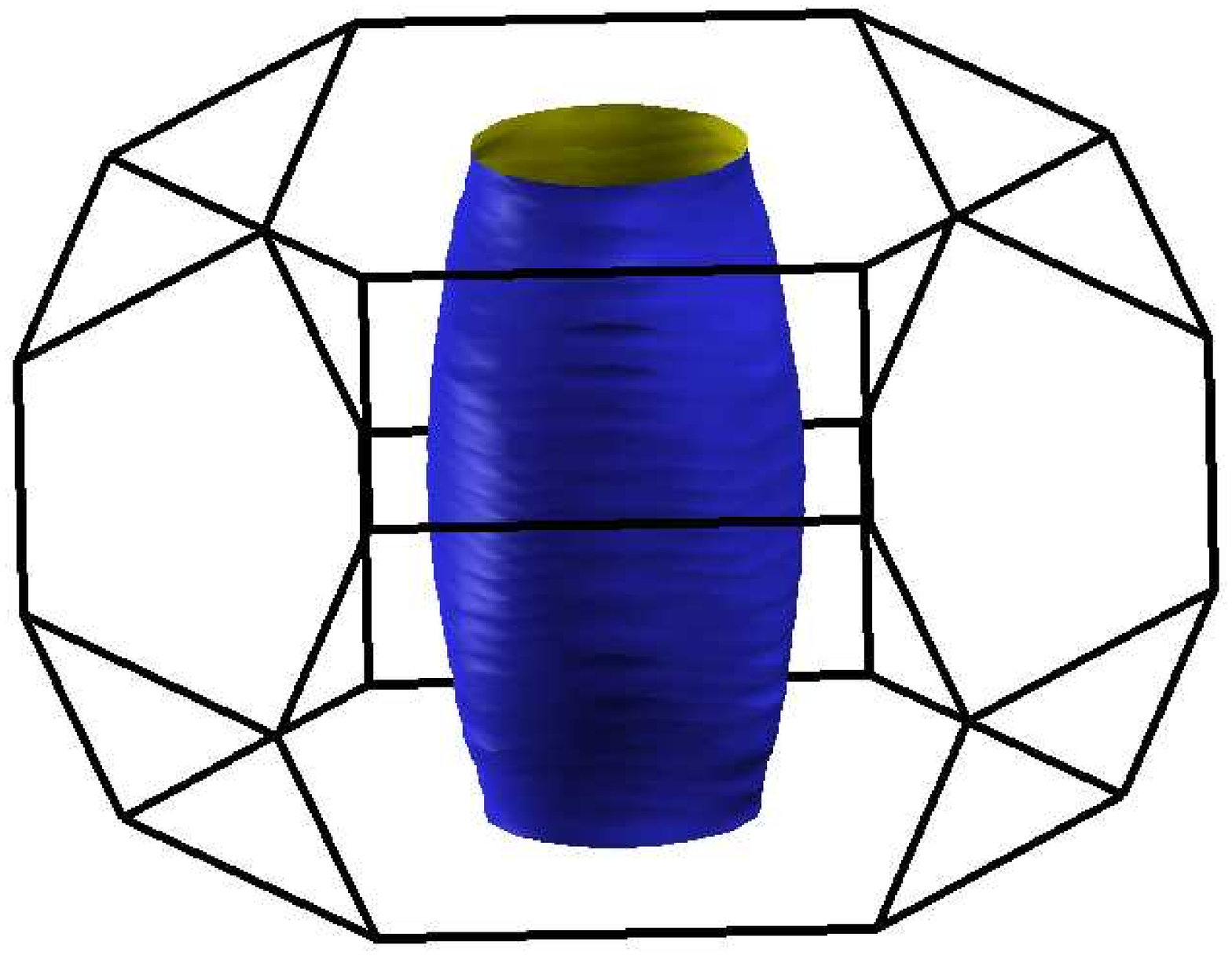}}
{\includegraphics[width=2.75cm]{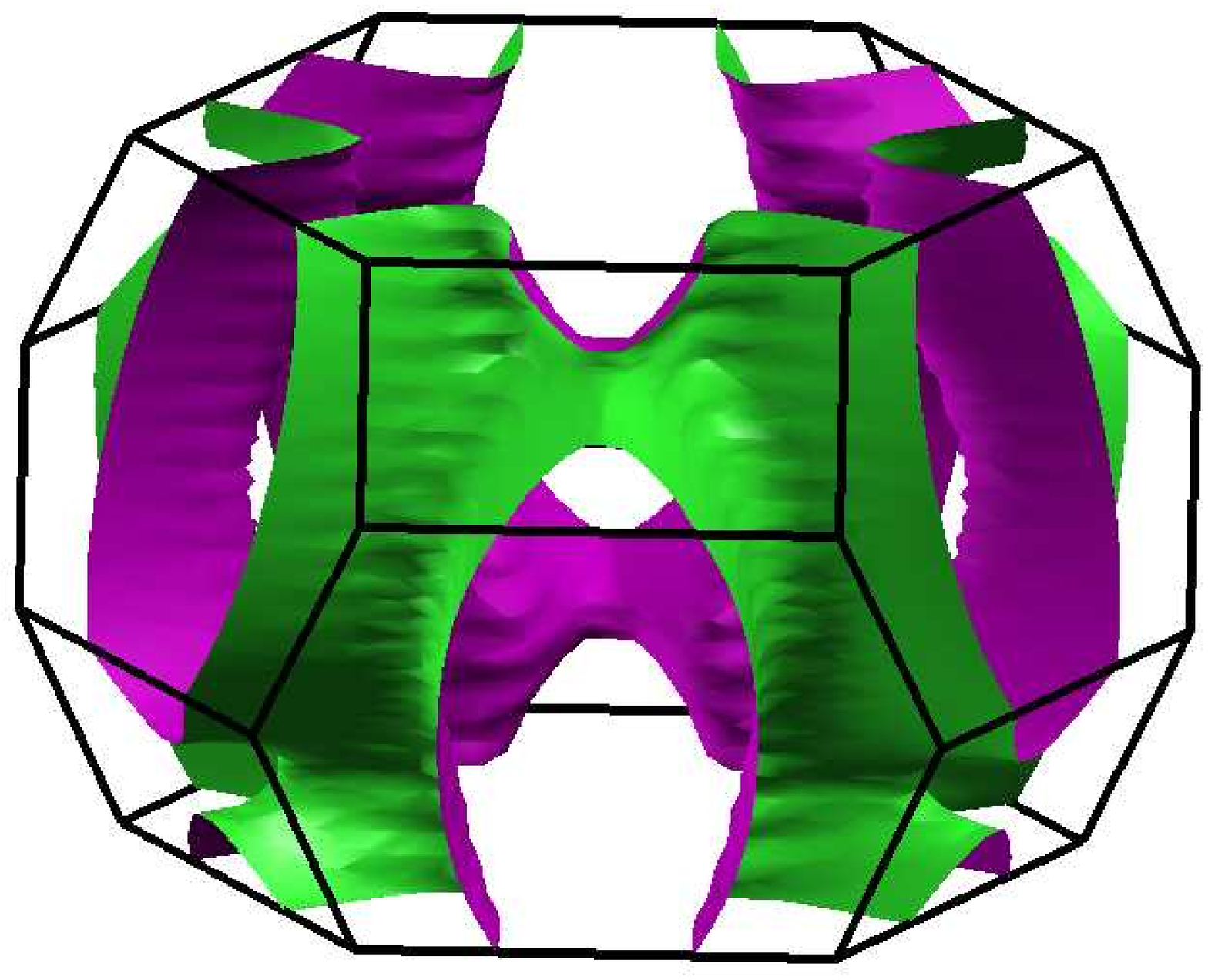}%
\includegraphics[width=2.75cm]{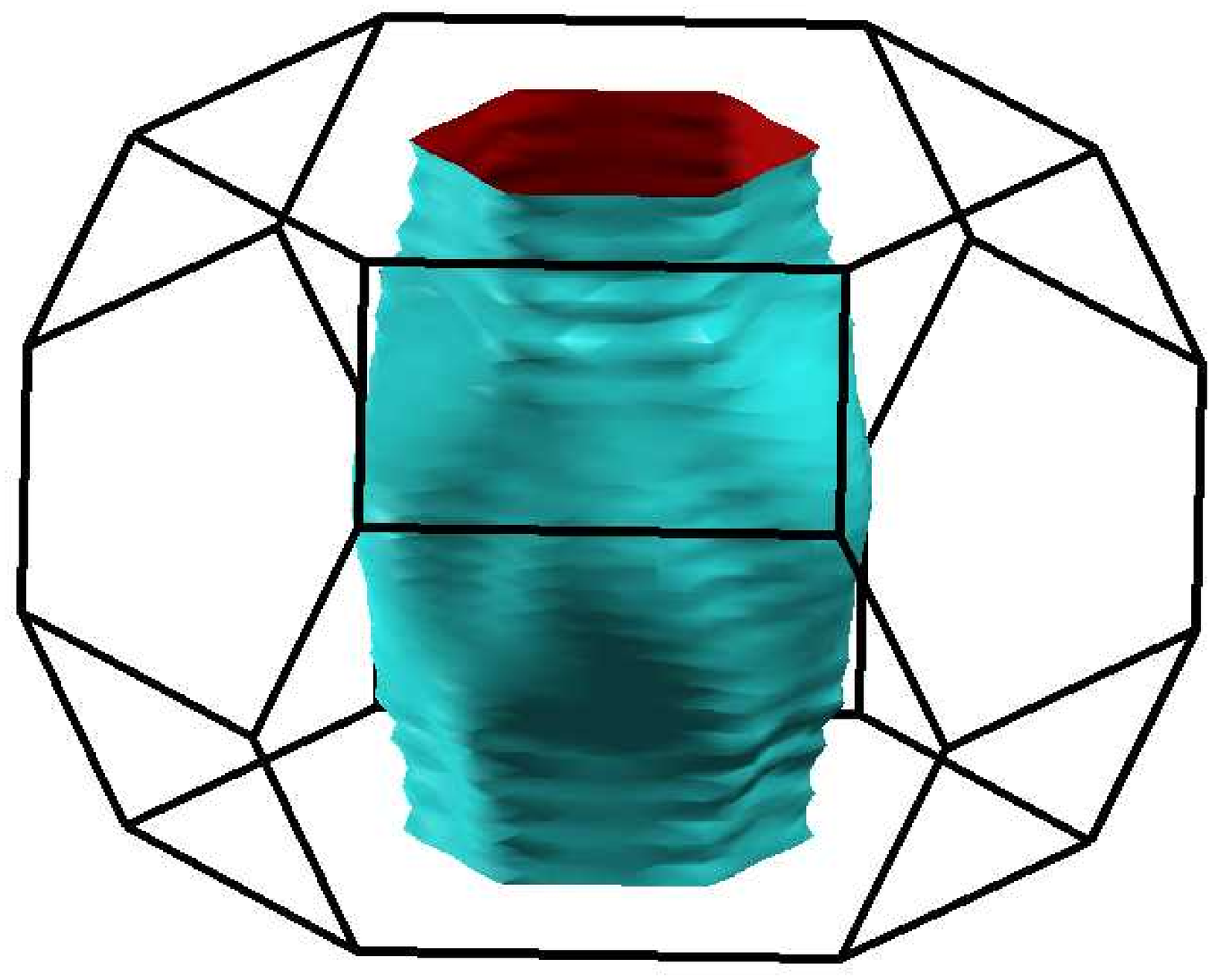}%
\includegraphics[width=2.75cm]{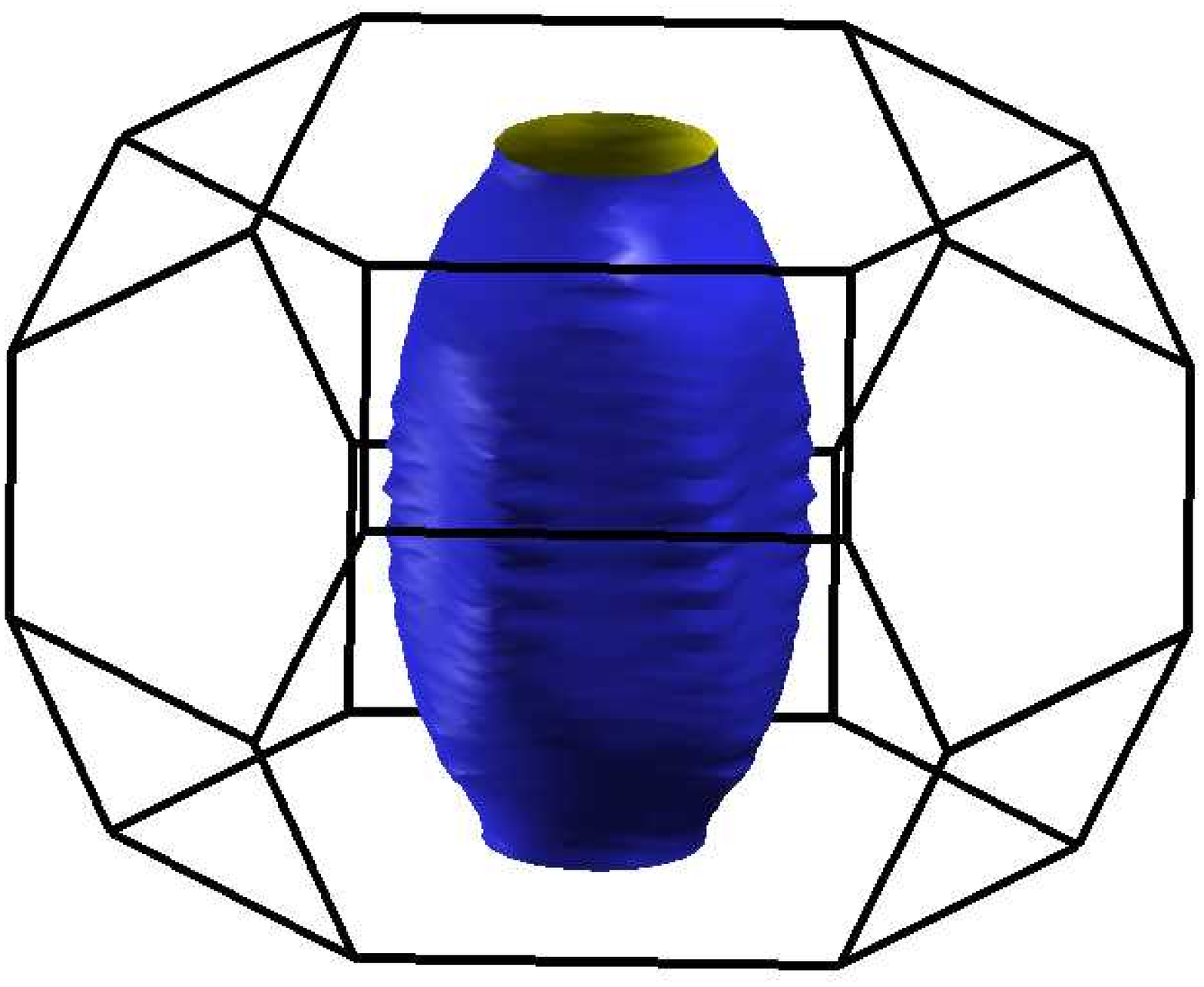}}
{\includegraphics[width=2.75cm]{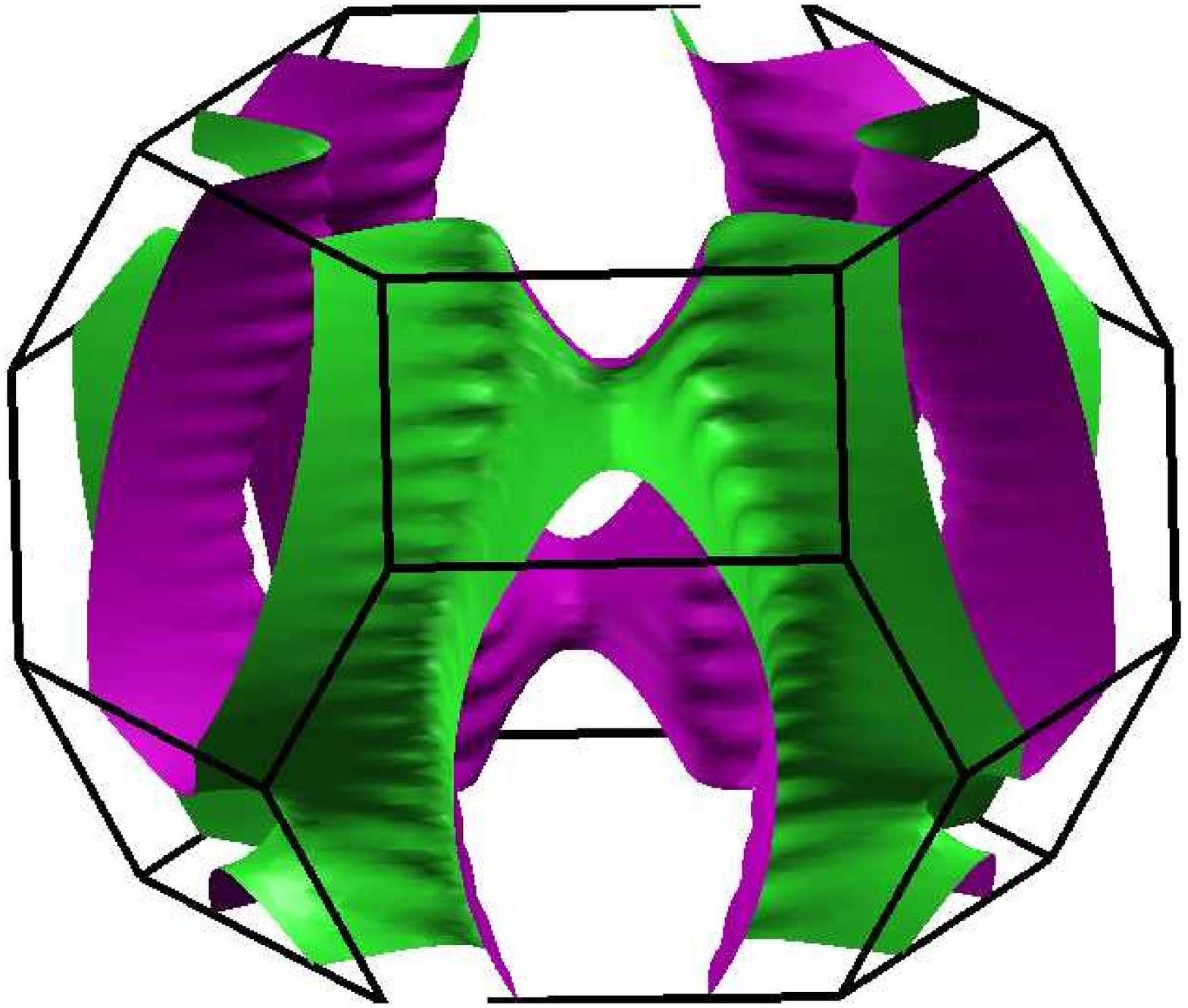}%
\includegraphics[width=2.75cm]{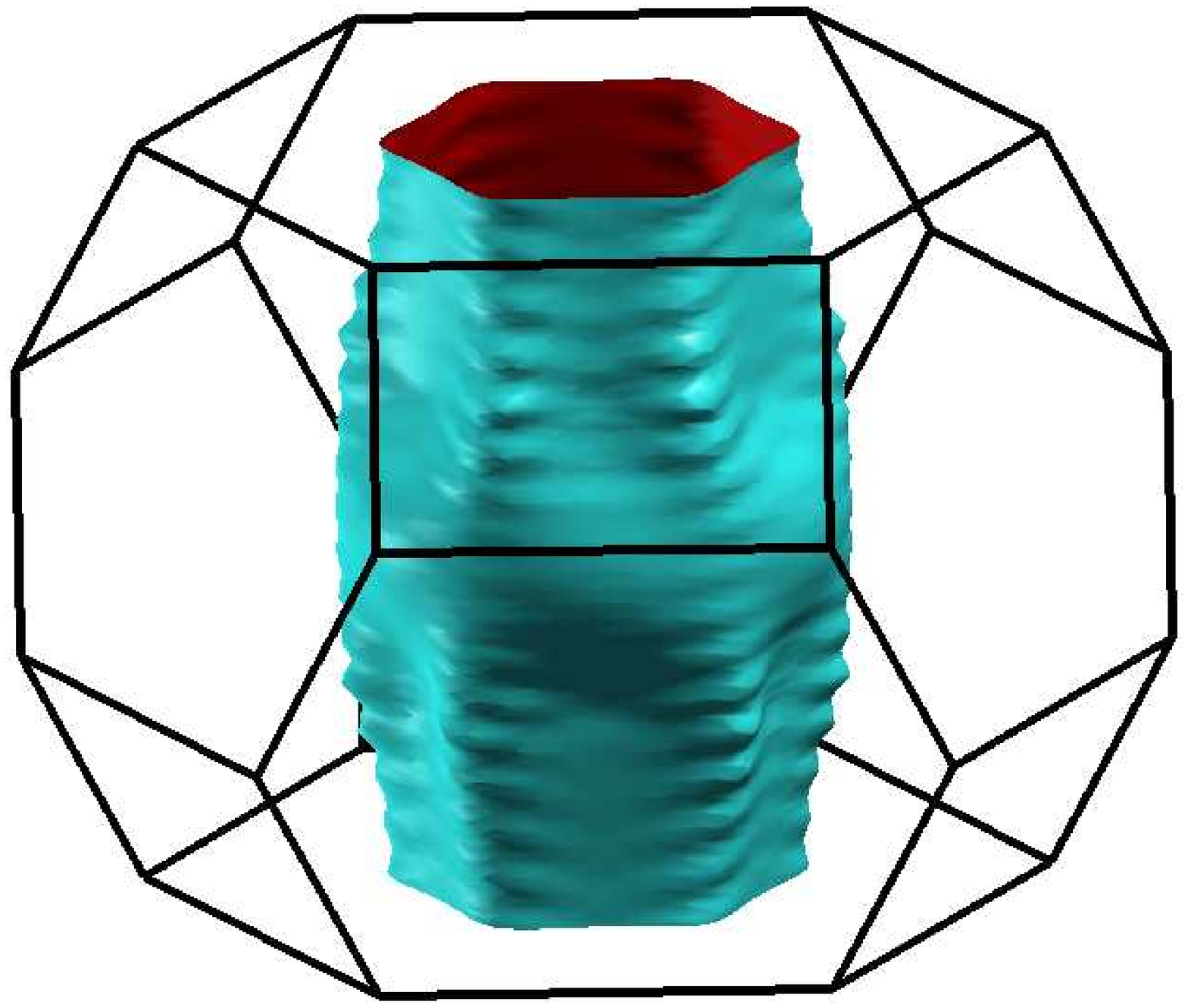}%
\includegraphics[width=2.75cm]{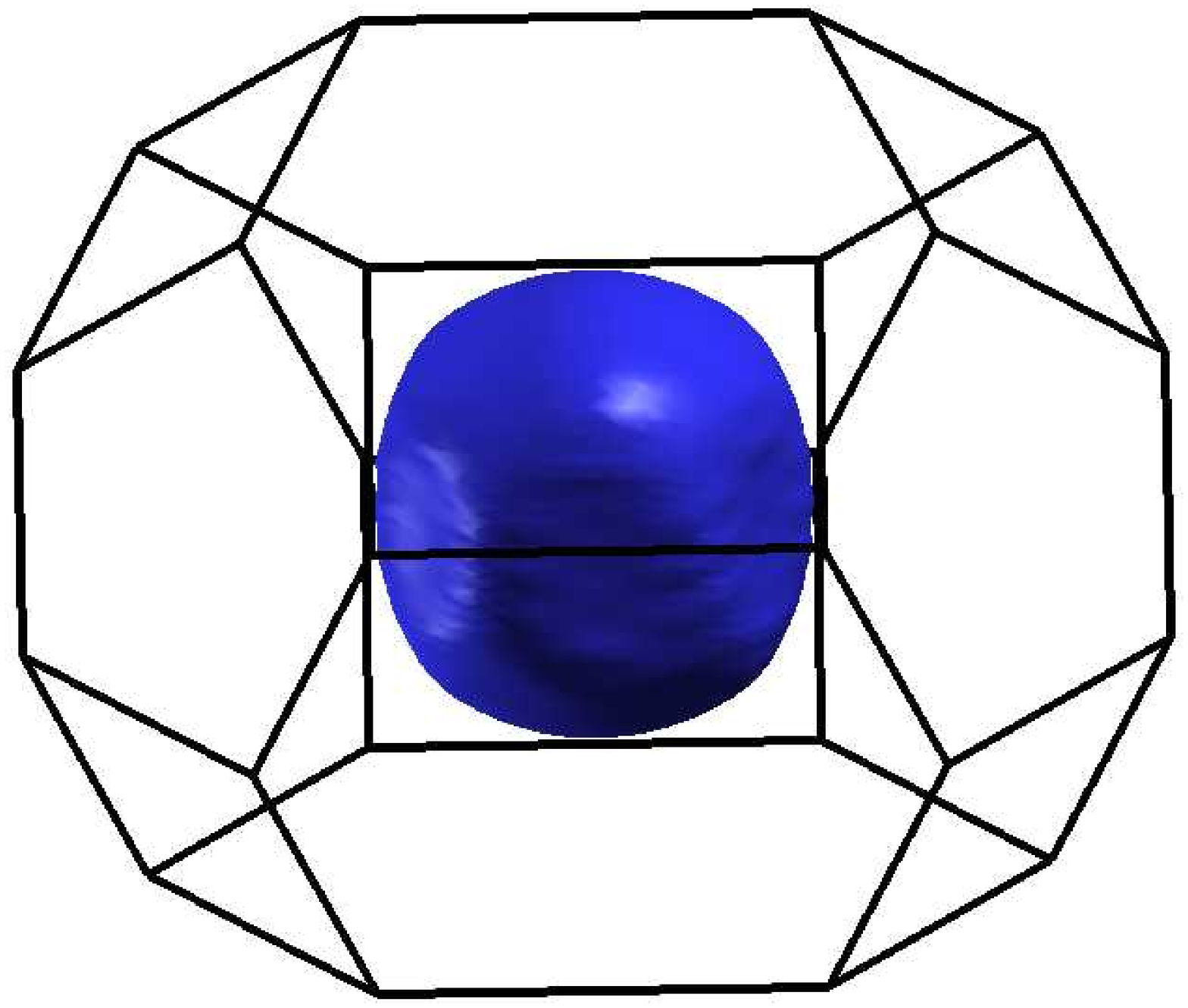}}
\caption{(Color online) Fermi surface sheets of BaC$_6$, SrC$_6$ and CaC$_6$ at
zero pressure. The sheets are plotted in order of increasing
band index.}
\label{fig:Fsurf}
\end{figure}

The three Fermi surface sheets of BaC$_6$, SrC$_6$ and CaC$_6$ are
shown in fig. \ref{fig:Fsurf}. The sheets have been identified 
by the corresponding band index. 
As c is reduced passing from BaC$_6$ to CaC$_6$ the main difference
is the change in the third sheet of the Fermi surface which
is indeed a warped cylinder in BaC$_6$ and it  becomes spherical in 
CaC$_6$. This is best seen analyzing the band structure in the $\Gamma L$
direction (namely the $k_z$ direction). In BaC$_6$ and SrC$_6$ there
are no bands crossing E$_f$ along $\Gamma L$, while in CaC$_6$ 
the Ca-originated band crosses E$_f$.

\section{Phonon dispersion}\label{sec:phonon}

The phonon dispersion of all the considered GICs can be divided in three
regions. The high energy region is composed by in-plane Carbon-vibration
( C$_{xy}$ ),
the intermediate energy region by out-of-plane Carbon vibration
( C$_{z}$ ),
and the low energy region is mostly composed intercalant vibration
( I$_{xy}$ for in-plane and I$_{z}$ for out-of-plane, with I=Ba,Sr,Ca,Mg),
as it was shown for CaC$_6$\cite{Calandra}. Since the high energy branches
($\omega_{{\bf q}\nu} > 100$meV ) are weakly affected by the 
type of intercalant, for the sake of clarity, in some case we only plot branches
in the low energy region of the spectra. These branches are those
having the greatest contribution to the electron-phonon coupling.

\begin{figure}[h]
\includegraphics[width=9.0cm]{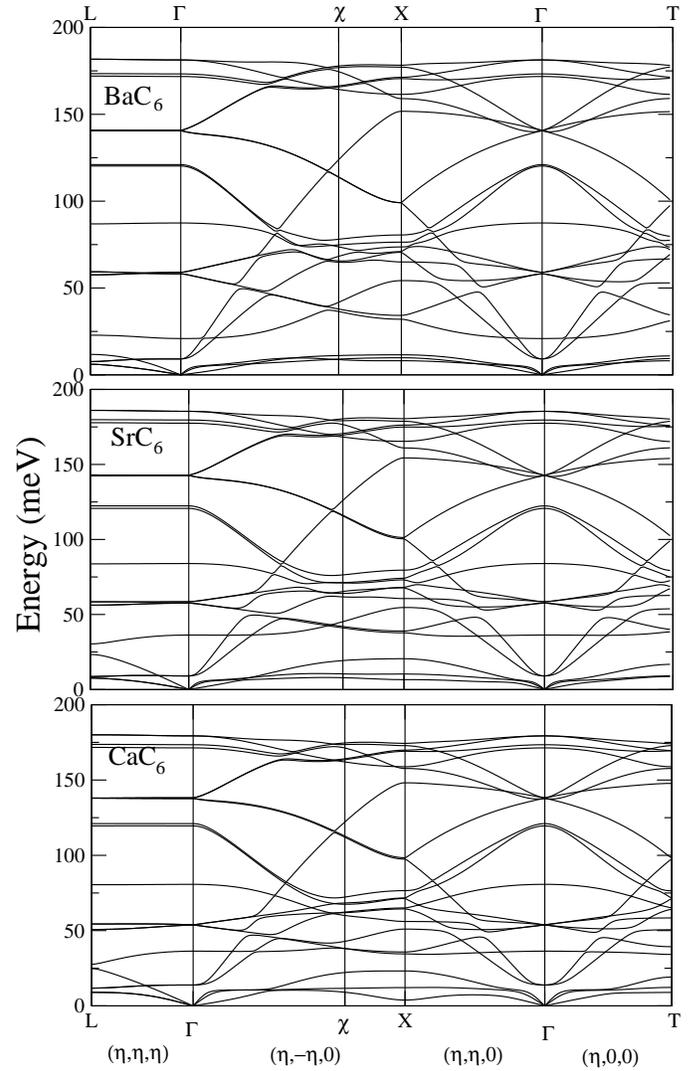}%
\caption{Phonon dispersions of
BaC$_6$, SrC$_6$ and CaC$_6$ at 0 GPa. }
\label{fig:PhononsP0large}
\end{figure}

\begin{figure}[h]
\includegraphics[width=9.0cm]{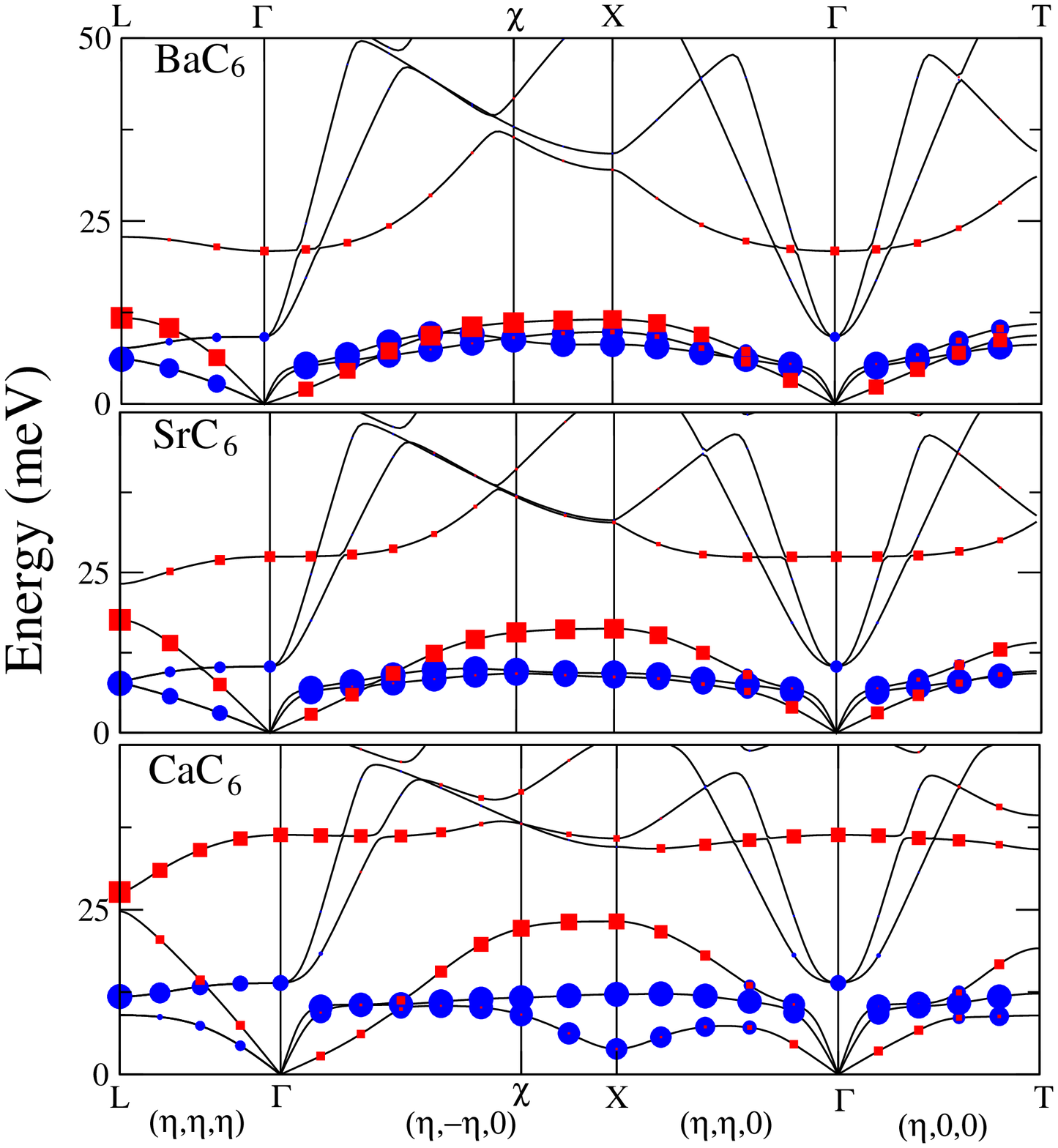}%
\caption{(Color online) Low energy region of the phonon dispersions of
BaC$_6$, SrC$_6$ and CaC$_6$ at 0 GPa. The size of the
blue dots (red squares) represents the amount of I$_{xy}$ (I$_{z}$),
where ${\rm I=Ba,Sr,Ca}$.
vibrations.}
\label{fig:PhononsP0}
\end{figure}

The zero-pressure phonon dispersion of BaC$_6$, SrC$_6$ and CaC$_6$
are illustrated in figs. \ref{fig:PhononsP0large} and \ref{fig:PhononsP0}.
As Z is reduced, and consequently c is reduced, the lowest energy branch in the
$\Gamma X$ direction (which corresponds to in-plane momenta with $k_z=0$) is
softened. This is evident in CaC$_6$ where an incipient Kohn anomaly
is found at X.  
While the reduction of c or Z softens the I$_{xy}$, the 
intercalant I$_z$ out-of-plane vibrations are hardened. 

This effect is also apparent from the analysis of the phonon dispersion of SrC$_6$ and
CaC$_6$ under pressure. In SrC$_6$ the lowest branch at 16 GPa in the $\Gamma X$ 
direction is almost imaginary, 
meaning that at this pressure the system is close to a structural
instability\cite{footnoteSrC6}.
This instability is driven by the reduction in the c-lattice spacing.
\begin{figure}[t]
\includegraphics[height=8.5cm]{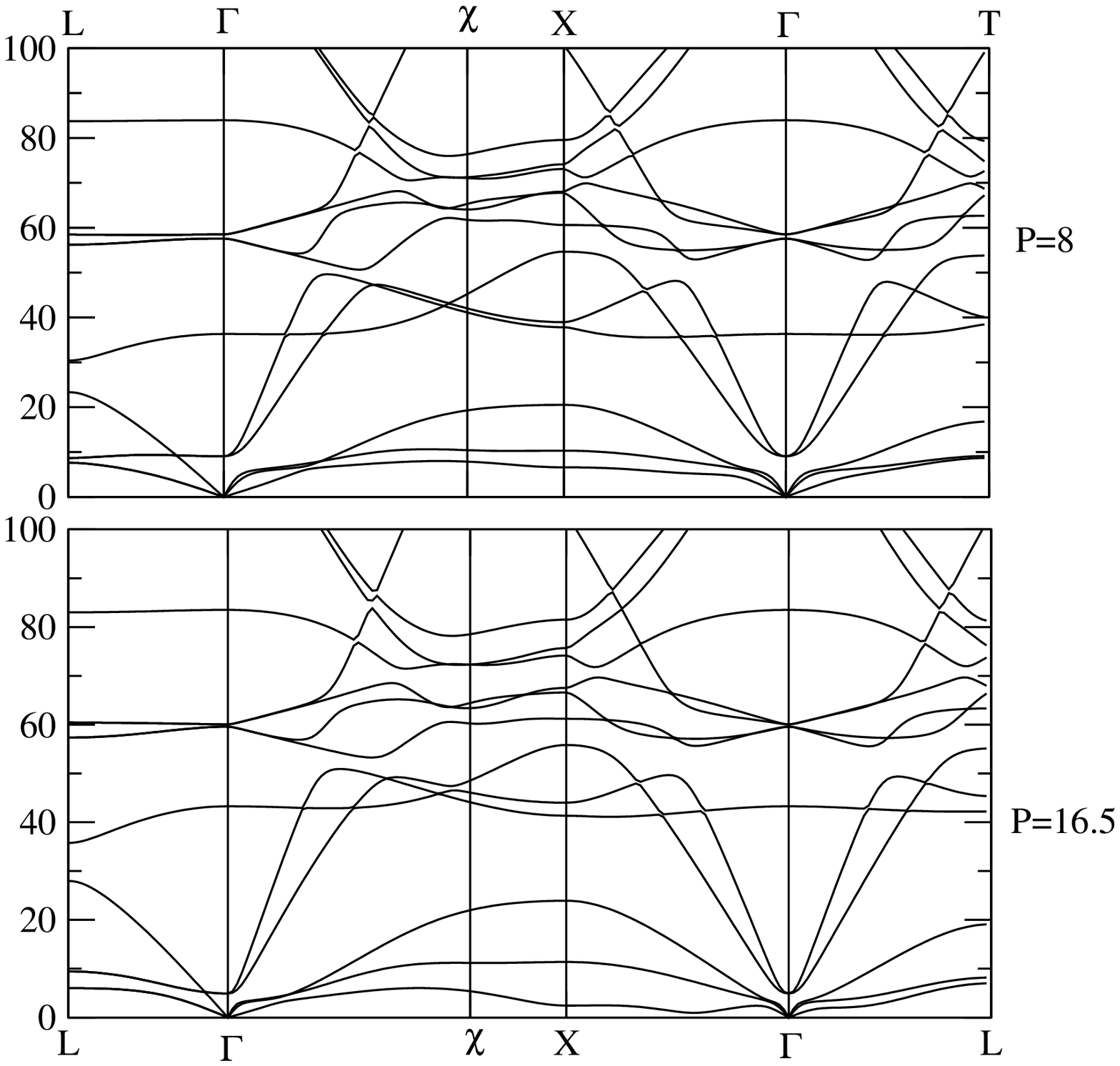}%
\caption{Phonon dispersions of SrC$_6$ at 8 and 16.5 GPa.}
\label{fig:PhononsSrP}
\end{figure}
In MgC$_6$, at the theoretical lattice parameters \ref{tab:structures}, we find
imaginary phonon frequencies at X, so that the system is probably unstable.
As a consequence we did not calculate the full phonon dispersion.

\section{Superconducting properties}\label{sec:SC}

The superconducting properties of Alkali-Earths GICs can be understood calculating
the electron-phonon coupling $\lambda_{{\bf q}\nu}$ 
for a phonon mode $\nu$ with momentum ${\bf q}$:
\begin{equation}\label{eq:elph}
\lambda_{{\bf q}\nu} = \frac{4}{\omega_{{\bf q}\nu}N(0) N_{k}} \sum_{{\bf k},n,m} 
|g_{{\bf k}n,{\bf k+q}m}^{\nu}|^2 \delta(\epsilon_{{\bf k}n}) \delta(\epsilon_{{\bf k+q}m})
\end{equation}
where the sum is over the Brillouin Zone.
The matrix element is
$g_{{\bf k}n,{\bf k+q}m}^{\nu}= \langle {\bf k}n|\delta V/\delta u_{{\bf q}\nu} |{\bf k+q} m\rangle /\sqrt{2 \omega_{{\bf q}\nu}}$,
where $u_{{\bf q}\nu}$ is the amplitude of the displacement of the phonon 
and $V$ is the Kohn-Sham potential.
The average electron-phonon coupling is  
$\lambda=\sum_{{\bf q}\nu} \lambda_{{\bf q}\nu}/N_q $ and its value for the different compounds and for
different pressures is given in table \ref{tab:SCpressGICS}.

\begin{figure}[t]
\includegraphics[height=8.5cm]{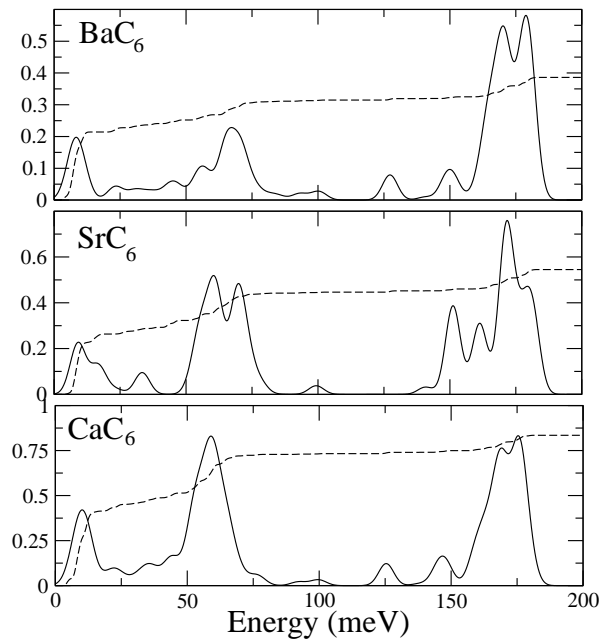}%
\caption{(a) Zero-pressure Eliashberg function, 
$\alpha^2F(\omega)$, (continuous line) and
integrated coupling, $\lambda(\omega)$ (dashed) for BaC$_6$, SrC$_6$ and CaC$_6$.}
\label{fig:alpha2fP0}
\end{figure}

The Eliashberg function
\begin{equation}
\alpha^2F(\omega)=\frac{1}{2 N_q}\sum_{{\bf q}\nu} \lambda_{{\bf q}\nu} \omega_{{\bf q}\nu} \delta(\omega-\omega_{{\bf q}\nu} )
\end{equation}
and the integral $\lambda(\omega)=2 \int_{-\infty}^{\omega} d\omega^{\prime} 
\alpha^2F(\omega^{\prime})/\omega^{\prime}$ are shown in fig. \ref{fig:alpha2fP0}
for BaC$_6$ and SrC$_6$. As for CaC$_6$ the largest contribution 
to $\lambda$ comes from modes below 75 meV.

An estimate of the different contributions of the in-plane and out-of-plane vibrations 
of the different atoms to $\lambda$ can be obtained from the relation
\begin{equation}\label{eq:trlambda}
\lambda=
\frac{1}{N_q}\sum_{\bf q}
\sum_{i\alpha j\beta} [{\bf G}_{\bf q}]_{i\alpha,j\beta} [{\bf C_q}^{-1}]_{j\beta,i\alpha}
\end{equation}
where $i,\alpha$ indexes indicate the displacement in the Cartesian direction $\alpha$
of the $i^{\rm th}$ atom,  
$[{\bf G_q}]_{i\alpha,j\beta}=\sum_{{\bf k},n,m}4 {\tilde g}_{i\alpha}^{*}{\tilde g}_{j\beta}
\delta(\epsilon_{{\bf k}n}) \delta(\epsilon_{{\bf k+q}m})/[N(0) N_{k}]$, and 
${\tilde g}_{i\alpha}=\langle {\bf k}n|\delta V/\delta x_{{\bf q} i\alpha} |{\bf k+q} m\rangle
 /\sqrt{2}$. 
The ${\bf C_q}$ matrix is the Fourier transform of the force constant matrix 
(the derivative of the forces respect to the atomic displacements).

Using eq. \ref{eq:trlambda} we decompose $\lambda$ restricting the summation 
over $i,\alpha$ and that over
$i,\beta$ on two sets of atoms and Cartesian directions. The sets are
C$_{xy}$, C$_{z}$, I$_{xy}$, and I$_z$, where I=\{Ba,Sr,Ca\}. 
The resulting $\bm{\lambda}$ matrix for
the different GICs and as a function of pressure are given in table
\ref{tab:lamdecpressGICS}.

\begin{table}[h]
\begin{tabular}{l|cccc}\hline
          &        &  C$_6$Ba & P=0       &          \\
          &  C$_{xy}$ & C$_{z}$ & Ba$_{xy}$ & Ba$_{z}$ \\ \hline
 C$_{xy}$ &  0.10  &   0.00 &  -0.01   &  0.00 \\	  
 C$_{z}$  & 0.00  &   0.10 &   0.01   & -0.01 \\	  
Ba$_{xy}$ & -0.01  &   0.01 &   0.12   & -0.00 \\	  
Ba$_{z}$  &  0.00  &  -0.01 &   0.00   &  0.07 \\ \hline
          &          &  C$_6$Ba & P=8       &          \\
          &   C$_{xy}$ & C$_{z}$ & Ba$_{xy}$ & Ba$_{z}$ \\ \hline
 C$_{xy}$ &  0.10  &    0.01   &  -0.00   &   0.00 \\	     
 C$_{z}$  &  0.01  &    0.11   &   0.02   &  -0.01 \\	     
Ba$_{xy}$ & -0.00  &    0.02   &   0.16   &  -0.00 \\	     
Ba$_{z}$  &  0.00  &   -0.01   &  -0.00   &   0.05 \\ \hline
          &          &  C$_6$Sr & P=0       &          \\
          &   C$_{xy}$ & C$_{z}$ & Sr$_{xy}$ & Sr$_{z}$ \\ \hline
 C$_{xy}$ &  0.12  &    0.00  &   -0.00  &    0.00 \\	     
 C$_{z}$  &  0.00  &    0.20  &    0.02  &   -0.00 \\	     
Sr$_{xy}$ & -0.00  &    0.02  &    0.16  &   -0.00 \\	     
Sr$_{z}$  &  0.00  &   -0.00  &   -0.00  &    0.05 \\ \hline
          &  C$_6$Sr & P=8       &          \\
          &   C$_{xy}$ & C$_{z}$ & Sr$_{xy}$ & Sr$_{z}$ \\ \hline
 C$_{xy}$ &  0.12  &    0.01  &   -0.00  &    0.00 \\	     
 C$_{z}$  &  0.01  &    0.24  &    0.03  &   -0.01 \\	     
Sr$_{xy}$ & -0.00  &    0.03  &    0.19  &   -0.00 \\	     
Sr$_{z}$  &  0.00  &   -0.01  &   -0.00  &    0.04 \\ \hline
          &          &  C$_6$Sr & P=16       &          \\
          &   C$_{xy}$ & C$_{z}$ & Sr$_{xy}$ & Sr$_{z}$ \\ \hline
 C$_{xy}$ & 0.13  &    0.03   &  -0.01  &    0.00 \\	     
 C$_{z}$  & 0.03  &    0.65   &   0.21  &   -0.01 \\	     
Sr$_{xy}$ &-0.01  &    0.21   &   0.71  &   -0.04 \\	     
Sr$_{z}$  & 0.00  &   -0.01   &  -0.04  &    0.12 \\ \hline
          &          &  C$_6$Ca & P=0       &          \\
          &   C$_{xy}$ & C$_{z}$ & Ca$_{xy}$ & Ca$_{z}$ \\ \hline
 C$_{xy}$ &  0.12  &    0.00  &   -0.00 &     0.00 \\	    
 C$_{z}$  &  0.00  &    0.33  &    0.04 &    -0.01 \\	    
Ca$_{xy}$ & -0.00  &    0.04  &    0.27 &    -0.00 \\	    
Ca$_{z}$  &  0.00  &   -0.01  &   -0.00 &     0.06 \\ \hline
 &          &  C$_6$Ca & P=5       &          \\
          &   C$_{xy}$ & C$_{z}$ & Ca$_{xy}$ & Ca$_{z}$ \\ \hline
 C$_{xy}$ &  .13  &    .01  &   -.00  &    .00 \\
 C$_{z}$  &  .01  &    .35  &    .03  &   -.01 \\
Ca$_{xy}$ & -.00  &    .03  &    .26  &   -.01 \\
Ca$_{z}$  &  .00  &   -.01  &   -.01  &    .05 \\ \hline

\end{tabular}
\caption{Decomposition of the electron-phonon coupling parameter into different
vibrational components for several GICs }
\label{tab:lamdecpressGICS}
\end{table}

Except for of SrC$_6$ at 16.5 GPa,
the off-diagonal matrix elements are negligible. 
In all the case the largest contributions to $\lambda$ comes
from C$_z$ and I$_{xy}$ phonon modes. The coupling to these modes
are enhanced as Z and c are reduced.
A similar enhancement of  C$_z$ and  I$_{xy}$ occurs under pressure.
When the pressure is too high, as in SrC$_6$ at P=16.5 GPa, the off-diagonal
matrix elements increase and the attribution to C$_z$ or to I$_{xy}$
becomes ill-defined. Note also that the C$_{xy}$ and the I$_z$ modes
are weakly affected by Z or c-axis reduction.

The critical superconducting temperature is estimated using the McMillan 
formula\cite{mcmillan}:
\begin{equation}
T_c = \frac{\langle \omega \rangle}{1.2}\, e^{ - \frac{1.04 (1+\lambda)}{\lambda-\mu^* (1+0.62\lambda)}}
\label{eq:mcmillan}
\end{equation}
where $\mu^*$ is the screened Coulomb pseudopotential and
\begin{equation} 
\langle\omega\rangle = e^{\frac{2}{\lambda}\int_{0}^{+\infty} 
\alpha^2F(\omega)\log(\omega)/\omega\,d\omega }
\end{equation}
the phonon frequencies logarithmic average. Results for $\langle\omega\rangle$ and
for T$_c$ are shown in table \ref{tab:SCpressGICS} using
$\mu^{*}=0.14$. For BaC$_6$ and SrC$_6$, 
T$_c$ increases as the interlayer spacing is decreased. For SrC$_6$ at 16.5
GPa the the result of McMillan formula is not 
correct. Indeed close to the transition 
the increase in $\lambda$ is only given by the softening of the 
Sr$_{xy}$ vibration close to X. In McMillan formula 
the limit $\omega_{{\bf q}\nu} \to 0$ implies  
T$_c \to 0$ as $\langle \omega \rangle$ since
$\lambda_{{\bf q}\nu} \to \infty$ as $1/\omega_{{\bf q}\nu}^{2}$.
Unfortunately the use of McMillan formula in this limit 
is not correct since it deviates
from the Migdal-Eliashberg results(ME), as known from
the papers of McMillan\cite{mcmillan} and of Allen and Dynes
\cite{AllenandDynes}(see fig. 1 in \cite{AllenandDynes}).  
For this reason the calculation of T$_c$ for SrC$_6$ at P=16.5 GPa
should be considered not reliable (and as a consequence it is not
given in table \ref{tab:SCpressGICS}).

\begin{table}[h]
\begin{tabular}{lccc}\hline
          &  C$_6$Ba &   &     \\
P(GPa) & $\lambda $ &  $\omega_{\rm log}$ (meV) & T$_c$($\mu^*$=0.14) K \\ \hline 
   0     &   0.38    &  22.44                  &    .23            \\ 
   8     &   0.45    &  20.63                  &    .81            \\ \hline
          &  C$_6$Sr &                                      \\
P(GPa) & $\lambda$ &  $\omega_{\rm log}$(meV) & T$_c$($\mu^*$=0.14) K \\ \hline 
   0     &   0.54    &  28.38                  &   3.03   \\
   8     &   0.65    &  23.40                  &   5.17   \\
  16.5   &   1.97    &   3.67                  &   {\rm nonsense}   \\ \hline
         &  C$_6$Ca &                \\
P(GPa) & $\lambda$ &  $\omega_{\rm log}$(meV) & T$_c$($\mu^*$=0.14) K \\ \hline 
   0     &   0.83    &  24.70                  &  11.03  \\
   5     &   0.845   &  24.6                   &  11.40       \\ \hline
\end{tabular}
\caption{Superconducting parameters of several GICs under pressure}
\label{tab:SCpressGICS}
\end{table}

\section{ ${\rm CaC}_6$ under pressure}\label{sec:CaC6P}

The study of CaC$_6$ under pressure needs particular emphasis due to the interest
motivated by the work of Gauzzi {\it et al.} \cite{Gauzzi} showing a considerable 
increase of Tc which reaches its maximum of 15.1 K at 7-8 GPa. The increase of T$_c$
has been confirmed by other two successive works measuring T$_c$ in a much smaller
range of pressure \cite{Kim1,Smith}. At 8GPa the system seems to undergo a phase 
transition towards a different structure with a lower superconducting T$_c$.

\begin{figure*}[]
\includegraphics[height=7.0cm]{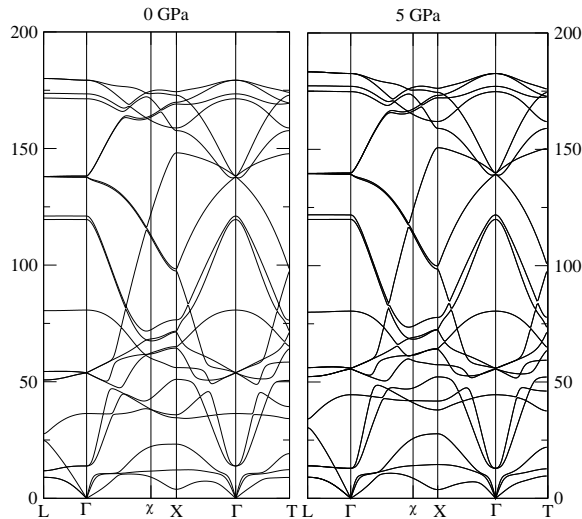}
\caption{Comparison between the phonon dispersion of CaC$_6$ at
0 GPa and at 5 GPa}.
\label{fig:breanchieCaC60-5Gpa}
\end{figure*}

\begin{figure*}[]
\includegraphics[height=7.0cm]{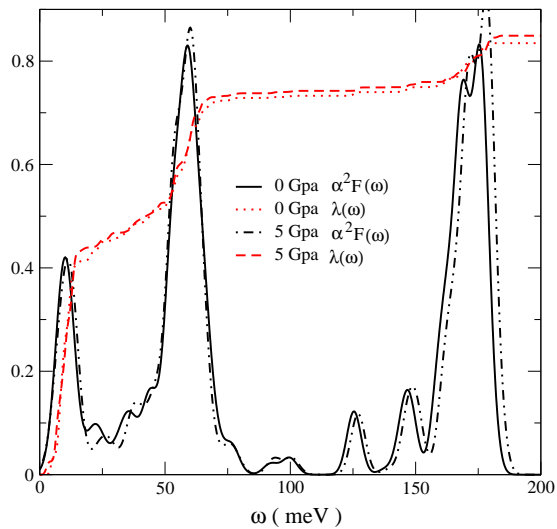}
\caption{(Color online) Comparison between $\alpha^2F(\omega)$ and $\lambda(\omega)$
at 0 GPa and 5 GPa for  CaC$_6$}.
\label{fig:alpha2fCaP}
\end{figure*}

To study the behavior of T$_c$ under pressure we calculate the phonon dispersion
and the electron-phonon coupling at 5GPa,
for which a T$_c=14 K$ was detected in experiments. The comparison between
the phonon dispersions at 0GPa and 5GPa is shown in fig. \ref{fig:breanchieCaC60-5Gpa}.
As can be seen, while the lowest Ca phonon mode is slightly softened (mostly at X), 
the others Ca and C$_z$ phonon modes are hardened. However the overall change is fairly
small and the Eliashberg functions $\alpha^2F(\omega)$ at 0 and 5GPa have no 
sizable differences (see fig. \ref{fig:alpha2fCaP}).  
At higher pressures the phonon frequencies
become imaginary, similarly to what happens in SrC$_6$ at 16.5GPa.
In particular the largest softening occurs at X for the lowest Ca mode and it becomes
negative in the range of 7-10 GPa. The lowest phonon frequency at X is extremely 
dependent on the Hermite-Gaussian smearing used in the calculation 
and as a consequence much larger k-point mesh must be used to correctly identify 
the phase transition (using a 10X10X10 mesh and smearing 0.04 still gives results which
are not converged). However it is clear that at a sufficient high pressure
the system will become unstable and the lowest Ca phonon frequency at X
will approach zero.

More insight on the superconducting properties can be gained by 
computing $\omega_{\rm log}$ and $\lambda$ at 0 and 5 GPa.
In this range of pressures these parameters have very similar values
and consequently the critical temperature changes only slightly
(see table \ref{tab:SCpressGICS}), from 11.03 K at 0GPa to 11.40 K at 5GPa.

The calculation for T$_c$ under pressure are in stark disagreement 
with a recent theoretical calculation\cite{Kim1} 
obtained using the same method (DFT) and the same code.

The disagreement with the  calculation in ref. \cite{Kim1} can be  
explained by the following.
In ref. \cite{Kim1} the electron-phonon coupling is calculated close to
the structural transition (at 10GPa) where the Ca phonon frequency strongly softens.
Just before the structural transition, $\omega_{\rm Ca}\mapsto 0$
and consequently $\lambda_{\rm Ca}\mapsto \infty$. 
In this limit (close to the transition),{\it the behavior is highly 
non linear}. On the contrary in ref.\cite{Kim1}
the behavior is assumed linear between 0GPa and 10GPa
\cite{footnote_linear}.
Our calculation at 5.0 GPa giving essentially the same T$_c$ as
at 0 GPa shows that this is not the case. 

We remark that in the calculation of the electron-phonon coupling at 0 and 5GPa
it is crucial to use at least a 4$\times$4$\times$4 q-point mesh.
If the smaller $2\times 2\times 2$ mesh is used, the electron-phonon
coupling of the Ca modes is overestimated, since
the point at X (which is included in the mesh
and has a considerable electron-phonon coupling) 
has a too large weight. The consequent increase of
T$_c$ under pressure is much larger than in the $4\times 4\times 4\times$ mesh.

From the preceding discussion it follows that DFT calculations
give a non linear behavior of T$_c$ versus pressure, in 
stark disagreement with experiments. In particular
the increase of T$_c$ as a function of pressure is too weak
when compared with experiments.
In the next section we discuss what can be the origin of such
disagreement.

\section{Conclusions}

In this work we theoretically investigated the  
occurrence of superconductivity in graphite intercalated 
with Alkaline earths (in particular Ba,Sr,Ca,Mg). 
Most of these systems have been already synthesized
(BaC$_6$,SrC$_6$ and CaC$_6$), while MgC$_6$ has 
been proposed\cite{Mazin} as a good candidate for
superconductivity due to its light mass and 
small force constants, possibly leading
to large electron-phonon coupling.

We predict the critical temperatures of BaC$_6$ and SrC$_6$
to be 0.23 K and 3.02 K respectively. Moreover we also predict
a substantial increase in the critical temperature under pressure
for these two systems, namely at 8 GPa the critical estimated
temperatures for BaC$_6$ and SrC$_6$ are 0.81 K and 5.16 K
respectively. We hope that these predictions can stimulate additional
experimental work so that the nature of the pairing in GICs can be 
further elucidated. Moreover it would be important to judge the 
reliability of DFT in calculating the superconducting properties of 
different GICs. Indeed, while superconducting gap and specific heat measurements
are in very good agreement with DFT calculations the only 
available measure of the Ca isotope effect \cite{Hinks} in 
CaC$_6$ is in stark disagreement with DFT predictions 
\cite{Calandra}. The claim that the isotope effect coefficient
of Ca is 0.5, if verified by the corresponding measurement of the C 
isotope effect, would open new perspectives. 

We have also shown that MgC$_6$ is energetically unstable against phase
separation in Mg and graphite. Moreover, even assuming that it can be synthesized
using pressure methods, the minimized structure at zero pressure obtained using 
DFT has imaginary phonon frequencies at the point X. Nonetheless it is interesting
to note that the softening at the point X occurs when the interlayer spacing
is reduced in CaC$_6$ and is connected to a large electron-phonon coupling
in this point.
This information suggests that the T$_c$  of CaC$_6$ can probably be raised 
synthesizing Mg$_x$Ca$_{1-x}$C$_y$ alloys.  

Concerning CaC$_6$, a puzzling problem is the dependence of the
critical superconducting temperature upon pressure. Experimentally T$_c$
increases as a function of pressure, but as we have shown in this work,
T$_c$ increases much faster than what DFT calculations predict. 
Further work is necessary to explain the origin of this
disagreement, mainly on the experimental side.
Indeed diffraction data as a function of pressure are absolutely
necessary. If there is a phase transformation upon pressure, the structure minimized 
with DFT at finite pressure would not be correct.
The fact that the Tc versus pressure curves are very smooth
seems to exclude an abrupt transition. A possibility is that a 
staging transition occurs in CaC$_6$ under pressure. This transition could
be continuous, meaning that the staging occurs progressively in 
the sample. It is indeed well known that GICs are extremely
sensitive to these kind of transitions occurring isothermally 
under very modest pressures. For example
KC$_{24}$, which is a stage 2 GIC, starts a staging transition
at 2.5 Kbar versus a stage 3 structure. The transition is continuous and
completely achieved at $\approx$ 6.5 Kbar\cite{Clarke}. 
Similar transitions have been
reported in ref. \cite{Wada} for MC$_8$ and MC$_{12n}$ (n=2,3,4) with M=Rb, Cs.
If such a transition occurs in CaC$_6$ then it would not necessarily show up
in the T$_c$ versus pressures curves, but it would explain the 
disagreement between experiment and theory.
Of course other more complicate explanations are possible, including
unconventional superconductivity.
In all the case, the origin of superconductivity in intercalated GICs is
not yet completely understood.

\section{Acknowledgements}

We acknowledge fruitful discussion with A. Gauzzi, G. Loupias, M. d'Astuto, 
N. Emery, C. Herold and L. Boeri. Calculations
were performed at the IDRIS supercomputing center (project 061202)



\begin{thebibliography}{99}
\bibitem{footnoteSrC6} The results for SrC$_6$ at 16.5 GPa should be considered
only qualitative since much larger q- and k-points mesh should be used. Moreover
very close to the structural instability anharmonic effects might become relevant.
\bibitem{footnote_linear} Even assuming a linear behavior, 
the theoretical Tc versus pressure slope is substantially smaller than the
experimental one.
\end{thebibliography}
\end{document}